\pdfoutput=1
\documentclass[aps,prd,twocolumn,superscriptaddress,preprintnumbers,nofootinbib]{revtex4}

\usepackage[pdftex]{graphicx}
\usepackage[latin1]{inputenc}
\usepackage{amsmath}
\usepackage{amsfonts}
\usepackage{amssymb}
\usepackage{rotating}
\usepackage{subfigure}
\usepackage{paralist} 
\usepackage{verbatim}
\usepackage{color}
\usepackage{soul} 
\usepackage{appendix}
\usepackage[flushleft]{threeparttable}

\usepackage{xcolor}

\begin{document}


\preprint{MIT-CTP 4505}


\title{Current Dark Matter Annihilation Constraints from CMB and Low-Redshift Data}

\author{Mathew S. Madhavacheril}
\affiliation{Stony Brook University, Stony Brook, New York, 11794, USA}

\author{Neelima Sehgal}
\affiliation{Stony Brook University, Stony Brook, New York, 11794, USA}

\author{Tracy R. Slatyer}
\affiliation{Massachusetts Institute of Technology, Cambridge, Massachusetts, 02139, USA}

\date{\today}

\begin{abstract}
Updated constraints on dark matter cross section and mass are presented combining CMB power spectrum measurements from Planck, WMAP9, ACT, and SPT  as well as several low-redshift datasets (BAO, HST, supernovae).  For the CMB datasets, we combine WMAP9 temperature and polarization data for $l \leq 431$ with Planck temperature data for $432 \leq l \leq 2500$, ACT and SPT data for $ l > 2500$, and Planck CMB four-point lensing measurements.  We allow for redshift-dependent energy deposition from dark matter annihilation by using a `universal' energy absorption curve.  We also include an updated treatment of the excitation, heating, and ionization energy fractions, and provide updated deposition efficiency factors ($f_\text{eff}$) for 41 different dark matter models.   Assuming perfect energy deposition ($f_\text{eff}=1$) and a thermal cross section, dark matter masses below 26 GeV are excluded at the $2\sigma$ level.   Assuming a more generic efficiency of $f_\text{eff}=0.2$, thermal dark matter masses below 5 GeV are disfavored at the $2\sigma$ level.  These limits are a factor of $\sim 2$ improvement over those from WMAP9 data alone. These current constraints probe, but do not exclude, dark matter as an explanation for reported anomalous indirect detection observations from AMS-02/PAMELA and the Fermi Gamma-ray Inner Galaxy data.  They also probe relevant models that would explain anomalous direct detection events from CDMS, CRESST, CoGeNT, and DAMA, as originating from a generic thermal WIMP.  Projected constraints from the full Planck release should improve the current limits by another factor of $\sim 2$, but will not definitely probe these signals. The proposed CMB Stage IV experiment will more decisively explore the relevant regions and improve upon the Planck constraints by another factor of $\sim 2$.
\end{abstract}

\pacs{}

\maketitle

\section{Introduction}

Non-baryonic matter is a crucial ingredient in our current understanding of the cosmological history of the Universe. A significant fraction of the energy density of the Universe is contended to consist of `dark matter' that interacts only very weakly (if at all) with ordinary matter.
 Dark matter is needed to explain numerous observations including gravitational lensing by clusters and galaxies, galaxy rotation curves, acoustic peaks in the power spectrum of the cosmic microwave background (CMB), and the growth of large-scale structure. However, all of the widely accepted evidence for dark matter is sensitive only to its gravitational effects, and the determination of its particle nature is an important open problem. Current efforts to address this can broadly be divided into \begin{inparaenum}[(i)] \item indirect detection experiments that aim to detect the products of dark matter annihilation or decay, \item direct detection experiments that attempt to detect dark matter particles via their recoil off heavy nuclei, and \item collider experiments where dark matter particles are hoped to be identified in the products of high-energy collisions.\end{inparaenum}

One particular indirect detection method is to observe the effect of dark matter annihilation early in the history of the Universe ($1400>z>100$) on the CMB temperature and polarization anisotropies \cite{Padmanabhan2005,Galli2009,Slatyer2009,Galli2011,Finkbeiner2012,Cline2013,Diamanti2013,Galli2013,LopezHonorez2013,Weniger2013,Natarajan2012}. If dark-matter particles self-annihilate at a sufficient rate, the expected signal would be directly sensitive to the thermally averaged cross section $\langle\sigma v\rangle$ of the dark matter particles
in this epoch, the mass $M_{\chi}$ of the annihilating particle, and the particular annihilation channel. An advantage of this indirect detection method over 
more local probes is that it is free of astrophysical uncertainties such as the local dark matter distribution and the astrophysical background of high-energy particles.  In Section \ref{sec:background}, we review the physics behind the modification of the CMB power spectra by annihilating dark matter.  We also discuss the universal energy deposition curve and systematic corrections to it as in \cite{Galli2013}, and the leverage in multipole-space of the dark matter constraints.  Updated constraints including all available data are presented in Section \ref{sec:constraints}. In Section \ref{sec:models}, we discuss these results in light of recent data from other indirect and direct dark matter searches.

\section{Effect of Dark Matter Annihilation on the CMB}\label{sec:background}

The recombination history of the Universe could potentially be modified by dark matter particles annihilating into Standard Model particles, which in turn inject energy into the (pre-recombination) photon-baryon plasma and (post-recombination) gas and background radiation.  Previous authors \cite{Padmanabhan2005,Galli2009,Slatyer2009,Galli2011,Finkbeiner2012} have considered the effects of this energy injection, which broadly consist of \begin{inparaenum}[(i)] \item increased ionization of the gas, \item atomic excitation of the gas, and \item plasma/gas heating. \end{inparaenum} These processes in turn lead to an increase in the residual ionization fraction ($x_e$) and baryon temperature ($T_b$) after recombination. For rates of energy injection low enough that there is minimal shift in the positions of the first few peaks of the CMB temperature power spectrum, the primary effect of the energy injection is to broaden the surface of last scattering. This leads to an attenuation of the temperature and polarization power spectra that is most pronounced at small scales.  In addition, the positions of the TE and EE peaks shift, and the power of polarization fluctuations at large scales ($l<500$) increases as the thickness of the last scattering surface grows.  (See Figure 4 in \cite{Padmanabhan2005} for a depiction of this effect.)

The rate of energy deposition per volume is given by, 
\begin{equation}
\frac{dE}{dV\,dt}=\rho_c^2c^2\Omega^2_\text{DM}(1+z)^6p_{\text{ann}}(z)
\end{equation} 
\begin{equation}
p_{\text{ann}}(z)=f(z)\frac{\langle\sigma v\rangle}{M_\chi} 
\end{equation} where $\rho_c$ is the critical density of the Universe today, $\Omega_\text{DM}$ is the density of cold dark matter today, $\langle\sigma v\rangle$ is the thermally averaged cross section of self-annihilating dark matter, $M_\chi$ is the dark matter mass, and $f(z)$ is an $\mathcal{O}(1)$ redshift-dependent function that describes the fraction of energy that is absorbed by the CMB plasma. In this parametrization, $f(z)$ captures the redshift-dependence of the energy deposition not included in the $(1+z)^3$ evolution of the dark matter density. The exact functional form of $f(z)$ depends on the specific annihilation channel of dark matter -- however, as discussed in \cite{Finkbeiner2012} and in Section \ref{sec:unisys}, the first principal component formed from the $f(z)$ energy deposition curves of 41 representative dark matter models accounts for more than 99.9\% of the variance in the CMB power spectra that is not degenerate with other standard cosmological parameters.  The injected energy modifies the evolution of the ionization fraction, $x_e$, according to
\begin{equation}
\frac{dx_e}{dz}=\frac{1}{(1+z)H(z)}[R_s(z)-I_s(z)-I_X(z)]
\label{eq:recomb}
\end{equation} 
where $R_s(z)$ and $I_s(z)$ are the standard recombination and ionization rates, respectively, in the absence of dark matter annihilation, $I_X(z)$ is the modification to ionization due to dark matter annihilation, and $H(z)$ is the Hubble constant at redshift $z$. Standard recombination, as discussed in \cite{Peebles1968}, is described by
\begin{equation}
[R_s(z)-I_s(z)]=C\times [x^2_en_H\alpha_B-\beta_B(1-x_e)e^{-h_P\nu_{2s}/k_BT_b}]
\end{equation} 
where the $C$-factor is given by 

\begin{equation}
C=\frac{[1+K\Lambda_{2s1s}n_H(1-x_e)]}{[1+K\Lambda_{2s1s}n_H(1-x_e)+K\beta_Bn_H(1-x_e)]}
\end{equation}
Here, $n_H$ is the hydrogen number density, $T_b$ is the baryon gas temperature, $\alpha_B$ and $\beta_B$ are the effective recombination and photoionization rates respectively for $n \geq 2$, $\nu_{2s}$ is the change in frequency from the $2s$ level to the ground state, $\Lambda_{2s1s}$ is the decay rate of the metastable $2s$ level to $1s$, $K=\lambda^3_\alpha/(8\pi H(z))$, and $\lambda_\alpha$ is the wavelength of the Lyman-$\alpha$ transition from $n=2$ to $n=1$.
This C-factor is approximately the probability that a hydrogen atom in the excited $n=2$ state will decay by two-photon emission to the $n=1$ state before being photodissociated \cite{Peebles1968}. 

Several authors have considered adding generic terms to the recombination equations, denoted by 
\begin{equation}
I_X(z) = I_{Xi}(z) + I_{X\alpha}(z),
\end{equation} 
that account for additional ionization from the ground state and from the $n=2$ state after energy injection \cite{Bean2007,Galli2008, Galli2009}. Dark matter annihilation increases the ionization fraction through  \begin{inparaenum}[(i)] \item direct ionization of hydrogen atoms from the ground state ($I_{Xi}(z)$), and \item ionization from the $n=2$ state after hydrogen has been excited by Lyman-$\alpha$ photons produced by dark matter annihilation ($I_{X\alpha}(z)$).\end{inparaenum} 
Following \cite{Galli2011}, the rate of additional ionization from the ground state is given by 
\begin{equation}
I_{Xi}=\chi_i\frac{[dE/dV\,dt]}{n_H(z)E_i}
\label{Eq:ionize}
\end{equation} 
where $E_i$ is the average ionization energy per baryon (13.6 eV), and $\chi_i$ is the fraction of absorbed energy that goes directly into ionization. 

The term describing ionization from the $n=2$ state is given by 
\begin{equation}
I_{X\alpha}=(1-C)\chi_\alpha\frac{[dE/dV\,dt]}{n_H(z)E_\alpha}
\label{Eq:excite}
\end{equation} 
where $\chi_\alpha$ is the fraction of absorbed energy that goes into excitation, $E_\alpha$ is the difference in binding energy between the $n=1$  and $n=2$ levels (10.2 eV), and $(1-C)$ is the probability of not decaying to the $n=1$ state before being photoionized from the $n=2$ state.

In addition, the baryon temperature evolution is modified by the last term in 
\begin{multline}
(1+z)\frac{dT_b}{dz}=\frac{8\sigma_Ta_RT^4_\text{CMB}}{3m_ecH(z)}\frac{x_e}{1+f_{\text{He}}+x_e}(T_b-T_\text{CMB}) \\
+2T_b - \frac{2}{3k_BH(z)}\frac{K_h}{1+f_{\text{He}}+x_e}
\end{multline}
where $f_\text{He}$ is the Helium fraction and
\begin{equation}
K_h=\chi_h\frac{[dE/dV\,dt]}{n_H(z)}.
\label{Eq:temp}
\end{equation} 
Here, $\chi_h$ is the absorbed energy converted to heat.   The energy fractions ($\chi_i, \chi_\alpha,$ and $\chi_h$) are discussed further in Section \ref{sec:unisys}.

\subsection{Universal Energy Deposition Curve with Systematic Corrections}\label{sec:unisys}

Many earlier studies of the impact of DM annihilation on recombination (e.g. \cite{Padmanabhan2005,Galli2009,Galli2011,Finkbeiner2012,Hutsi2011,Natarajan2009,Giesen2012,LopezHonorez2013,Natarajan2012}) have used an approximate form for the energy fractions $\chi_i, \chi_\alpha,$ and $\chi_h$, derived from Monte Carlo studies by Shull and van Steenberg in 1985 \cite{Shull1985}, and following the approximate fit suggested in \cite{Chen2004}:
\begin{eqnarray}
 \chi_i&=&\chi_e=\frac{(1-x_\mathrm{H})}{3} \nonumber\\ 
\chi_h&=&\frac{1+2x_\mathrm{H}+f_\mathrm{He}(1+2x_\mathrm{He})}{3(1+f_\mathrm{He})}.
\label{chi_iah}
\end{eqnarray}
Here $\chi_i$ is the hydrogen ionization fraction, $\chi_e$ is the hydrogen excitation fraction, and $\chi_h$ is the heating fraction.  The Lyman-$\alpha$ contribution, $\chi_\alpha$, is some fraction of $\chi_e$.  Some past studies have taken $\chi_\alpha = 0$ to obtain conservative constraints, while others, including this work, set $\chi_\alpha = \chi_e$. The helium fraction $f_\mathrm{He}$ is given by $f_\mathrm{He}=Y_p/(4(1-Y_p))$, where $Y_p$ is the helium mass fraction.   The ratio of ionized hydrogen to total hydrogen is given by $x_\mathrm{H}$, and the ratio of ionized helium to total helium is given by $x_\mathrm{He}$.  In this work, we do not include ionization of helium due to dark matter annihilations since it has a negligible impact on the CMB power spectra \cite{Galli2013,Giesen2012}.

In reality, the dependence of the energy fractions on the background ionization fraction $x_\mathrm{H}$ is more complex than the simple linear dependence in Eq. \ref{chi_iah}. The energy fractions also possess a non-trivial dependence on the energy of the electron when it is ``deposited'' to the plasma (i.e. when its energy drops to the point where all subsequent cooling processes have timescales much faster than a Hubble time). In previous work (e.g. \cite{Slatyer2009}), ``deposited'' photons with energies above 13.6 eV were treated exactly as deposited electrons, under the presumption that such photons would quickly ionize the gas, producing a free electron. While this is true, it is important to also account for the energy absorbed in the ionization itself. The free electron produced by photoionization will then deposit its energy subject to the appropriate energy fractions.

In this work we take these effects into account following the method described in detail in \cite{Galli2013}; our results use the same set of assumptions as that paper's ``best estimate'' constraints. Electrons, positrons and photons injected by DM annihilation are tracked down to a deposition scale of 3 keV, taking the expansion of the universe into account, using an improved version of the code first described in \cite{Slatyer2009}. The spectra of photons and electrons below this energy are stored -- many of the energy-loss processes are discrete rather than continuous, and thus these spectra are not simply spikes at the deposition scale -- and then integrated over energy-dependent energy loss fractions computed by  Monte Carlo methods, following \cite{Valdes2008, Furlanetto2010,Valdes2010,Evoli2013}. This part of the code does not take redshifting into account, but at energies below 3 keV all cooling times are much faster than a Hubble time (with the notable exception of photons below 10.2 eV after the redshift of last scattering), so the expansion can be neglected. Energy losses to direct ionization, excitation, and heating by electrons and photons above the 3 keV threshold are calculated in the ``high-energy'' code (appropriate to energies above 3 keV) and added to the corresponding fractions. ``Continuum'' (below 10.2 eV) and Lyman-alpha photons produced by inverse Compton scattering (ICS) of electrons above 3 keV are likewise calculated in the high-energy code; for electrons below 3 keV, ICS quickly becomes subdominant to atomic energy loss processes. Ionizations on helium are taken into account following \cite{Galli2013}.

The primary difference between the results of this method and earlier approximations is that the correct treatment of ICS by non-relativistic electrons predicts greater energy transfer into continuum photons, which cannot subsequently induce ionizations or Lyman-alpha excitations; the effect can be regarded as a high-energy distortion to the CMB energy spectrum. Consequently, the fraction of power going into ionization, excitation, and heating of the gas is somewhat depressed. There is an exception at high redshifts, where accounting for the additional ionization from \emph{photon}-gas interactions (which was not done in e.g. \cite{Slatyer2009}, which treated low-energy electrons and photons as identical) can outweigh the reduced ionization from electron-gas interactions, since the latter is very small in any treatment (those electrons lose their energy dominantly to Coulomb heating, using either the approximate fractions or the more accurate ones).

We have computed the fraction of deposited energy going into ionization, $\chi_i$, which largely controls the constraints (the Lyman-alpha fraction, $\chi_\alpha$, has a small, albeit not negligible, effect \cite{Galli2013}), as a function of redshift, for each of the 41 annihilation channels described in \cite{Slatyer2009}. The calculations of the energy fractions in \cite{Galli2013} separately compute the ionization on helium; here we simply sum the total power into ionization on hydrogen and helium to obtain the $\chi_i$ fraction, since as mentioned previously, the effects of separating the helium fraction are small. For convenience, given the widespread use of the approximate fractions of Eq.~\ref{chi_iah} in the literature and in existing code, for each annihilation channel we can define a new ``effective $f(z)$ curve'', $f_\mathrm{sys}(z)$, which yields the correct power-into-ionization when multiplied by the \emph{approximate} value of $\chi_i$. That is,
\begin{equation} 
\chi_i^\mathrm{approx}(z) f_\mathrm{sys}(z) = \chi_i^\mathrm{updated}(z) f_\mathrm{old}(z), 
\end{equation}
where $\chi_i^\mathrm{approx}$ and $\chi_i^\mathrm{updated}$ are respectively the approximate (Eq. \ref{chi_iah}) and updated (following \cite{Galli2013}) energy fractions, and $f_\mathrm{old}(z)$ agrees with the results of \cite{Slatyer2009}. (Note that in some cases this definition can lead to a very large value of $f_\mathrm{sys}(z)$, much greater than 1, where $\chi_i^\mathrm{approx}(z) \ll \chi_i^\mathrm{updated}(z)$.) This curve should not generally be applied to compute the heating and Lyman-$\alpha$ components, in cases where they are important; it is designed to correctly normalize the power into ionization. However, since we expect the effect of additional ionizations to dominate over the modification due to excitations or heating, we use the same $f_\text{sys}(z)$ curve for the ionization, excitation, and heating terms. We checked that using the $f_\text{sys}(z)$ curve to multiply the ionization term and the old $f(z)$ curve for the excitation and heating terms makes no appreciable difference to the constraints obtained below.

\begin{figure}[t!h]
     \hspace{-0.7cm}\includegraphics[width=.53\textwidth,natwidth=800,natheight=600]{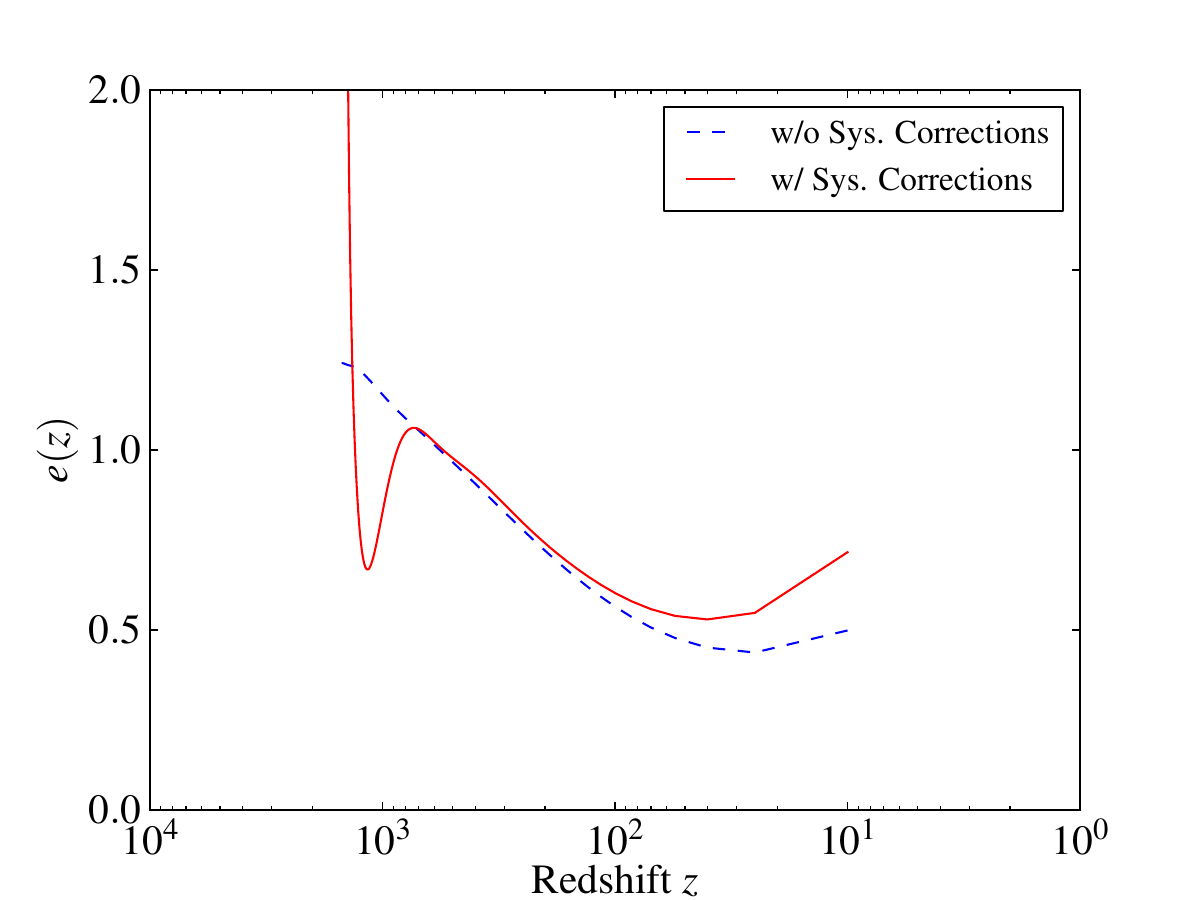}
     \caption{Universal energy deposition curve, $e(z)$, using approximations for the fraction of energy converted to heat, ionization, and excitation (dashed blue curve), and accounting for more accurate calculations of the energy fractions from \cite{Galli2013} (solid red curve).}
     \label{fig:ewimp}
\end{figure}

Having derived new individual $f_\mathrm{sys}(z)$ curves for a range of Standard Model final states, we can perform a principal component analysis using these curves as basis vectors, as described in detail in \cite{Finkbeiner2012}. The first principal component describes the direction in this space (of linear combinations of the $f_\mathrm{sys}(z)$ curves) which captures the greatest amount of the variance in the CMB power spectra -- in this case, over 99.9\%. Physically, the effects of the different annihilation channels on the CMB anisotropy spectra are very similar. 

We show in Figure \ref{fig:ewimp} the resulting first principal component as a function of redshift, which we refer to as the ``universal'' $e(z)$ curve. The overall normalization of the curve is arbitrary since it is precisely its amplitude that we wish to constrain, and hence a rescaling of $e(z)$ would be reflected in a proportional rescaling of the derived constraint on its coefficient. In order to fix the normalization, we adopt the convention used in \cite{Finkbeiner2012}, i.e., we fix the normalization such that if $p_{\text{ann}}(z)=\epsilon~e(z)$, the Fisher matrix constraint on $\epsilon$ is the same as that obtained for constant annihilation, $p_{\text{ann}}=\epsilon$ (with approximate energy fractions), for some choice of experimental parameters. The advantage of this choice is that constraints on the coefficient of $e(z)$ can be directly compared to previously derived constraints using constant $p_\text{ann}$. In this work, the Fisher matrix computation and principal component analysis were performed for a Planck-like experiment in the range $\ell < 6000$; we have verified that performing the  analysis  instead for a cosmic variance limited (CVL) experiment in this $\ell$ range changes the shape and normalization of the  $e(z)$ curve only at the sub-percent level. The principal components do not change appreciably when additional cosmological parameters that could be degenerate with the annihilation parameter are added. This is discussed in Appendix A5 of \cite{Finkbeiner2012}.

Note that this choice of normalization means that the $e(z)$ curve does not reflect the general reduction in amplitude of the $f_\mathrm{sys}(z)$ curves relative to the older $f(z)$ curves, arising from the fact that $\chi_i^\mathrm{updated}(z)$ is generally lower than $\chi_i^\mathrm{approx}(z)$. To the degree that the Fisher matrix approach is valid, we expect the constraint on the coefficient of the updated $e(z)$ curve to be identical to the corresponding bound for the older $e(z)$ curve presented in \cite{Finkbeiner2012}, since both should be equivalent to the constraint using constant $p_\mathrm{ann}$ and approximate energy fractions. However, constraints on specific \emph{models} will change.

To translate from constraints on the coefficient of the $e(z)$ curve to constraints on a specific model, one must extract the coefficient of the first principal component, when the $f_\mathrm{sys}(z)$ curve for that model is expanded in the basis of principal components. This is referred to in \cite{Finkbeiner2012} and \cite{Slatyer2012} as taking a ``dot product'', but there is a subtlety here in that the dot product must be taken in the space defined by the 41 $f_\mathrm{sys}(z)$ curves, not in the space of functions of $z$. In the Fisher matrix approach, this corresponds to taking the dot product between the (discretized) $f_\mathrm{sys}(z)$ curve for that particular model and the vector $(e)^T F$, where $e$ is the (discretized) universal $e(z)$ curve, and $F$ is the marginalized Fisher matrix describing the effect on the CMB of energy depositions localized in redshift (see \cite{Finkbeiner2012} for the precise construction). The dot product is normalized by dividing by the result where $f_\mathrm{sys}(z)$ is replaced with $e(z)$, to obtain an ``effective $f$'' value $f_\mathrm{eff,new}$:

\begin{equation} f_\mathrm{eff,new} = \frac{e(z) \cdot F \cdot f_\mathrm{sys}(z)}{e(z) \cdot F \cdot e(z)}.\end{equation}

Below we present constraints on the dimensionful parameter $\epsilon$, which we label as $p_\text{ann}$ in Table \ref{table:constraints} for ease of comparison with the constant $p_\text{ann}$ case and general familiarity with that variable. In order to obtain a constraint on $\langle \sigma v \rangle / M_\chi$  for a specific DM model, the bound on $p_\text{ann}$ should be divided by $f_\mathrm{eff,new}$ for that model since
\begin{equation}
p_\text{ann} = f_\mathrm{eff,new} \frac{\langle \sigma v \rangle}{M_\chi}.
\end{equation}
(By definition, if $f_\mathrm{sys}(z) = e(z)$, then $f_\mathrm{eff,new} = 1$; the derived constraint on $p_\text{ann}$ is exactly the constraint on $\langle \sigma v \rangle / M_\chi$ for such a model.) We have verified that this prescription accurately reproduces the constraints presented for individual leptonic annihilation channels in \cite{Galli2013}. The fact that the $f_\mathrm{sys}(z)$ curves are generally lower than the original $f(z)$ curves is reflected in lower $f_\mathrm{eff,new}$ values, and hence weaker constraints on $\langle \sigma v \rangle / M_\chi$.

In Table \ref{table:feffs}, we provide both the $f_\mathrm{eff,new}$ values computed using our new $f_\mathrm{sys}(z)$ curves, and the $f_\mathrm{eff}$ values computed using the old $f(z)$ curves from \cite{Slatyer2009}, but using the correct Fisher-matrix weighting described in the previous paragraph (these values were computed in an online supplement to \cite{Finkbeiner2012}, but the dot product was not properly weighted by the Fisher matrix, leading to few-percent deviations).

\subsection{Leverage in $ \ell$-space of Dark Matter Limits}\label{sec:scale}

\begin{figure}[t!]
     \hspace{-0.65cm}\includegraphics[width=.53\textwidth,natwidth=800,natheight=600]{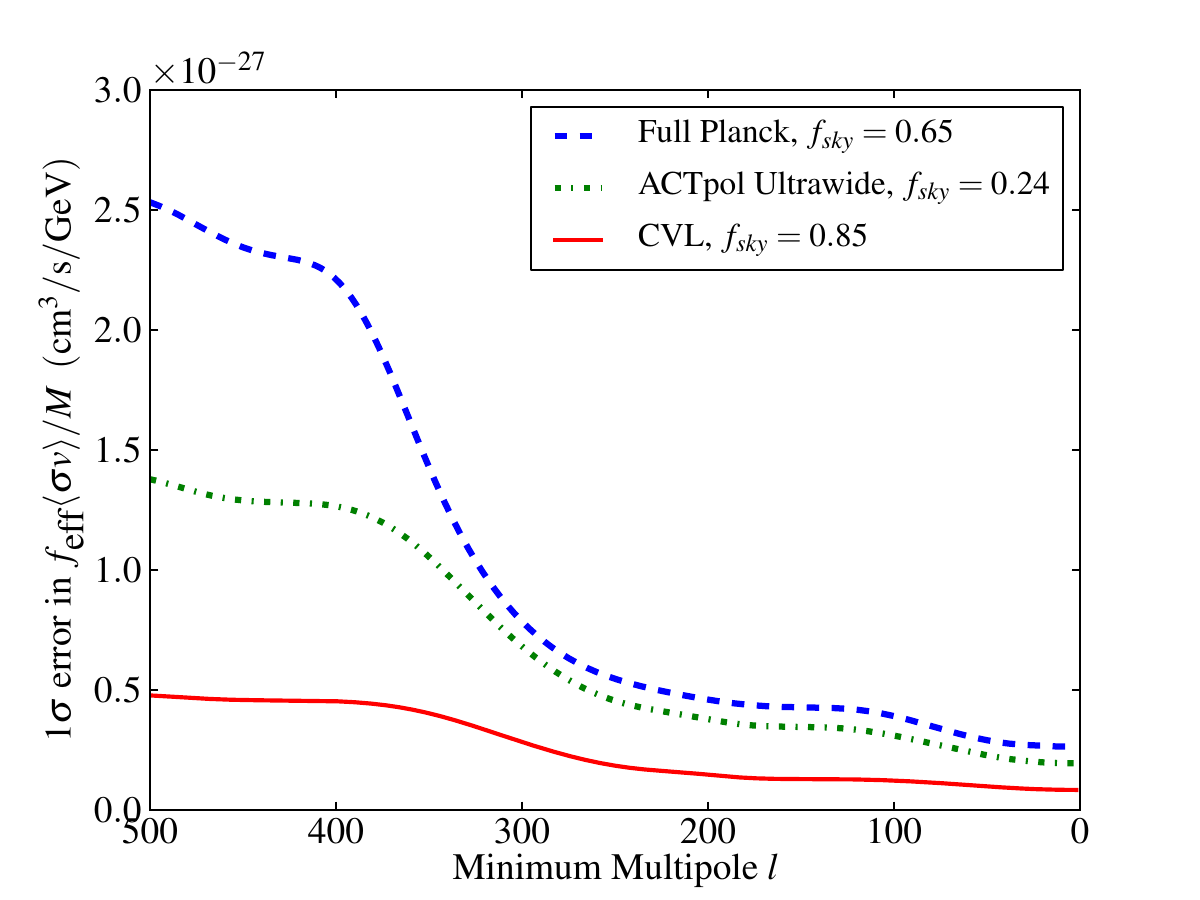}
     \caption{Fisher projected constraint obtained by including the range $500<l<5000$ and extending it cumulatively for each multipole below $l=500$. Experimental parameters are from Planck, an ACTpol-like experiment, and a cosmic variance limited experiment (see Table \ref{tab:specs}). Most of the leverage comes from $250<l<400$.}
     \label{fig:lowl}

      \hspace{-0.65cm}\includegraphics[width=.53\textwidth,natwidth=800,natheight=600]{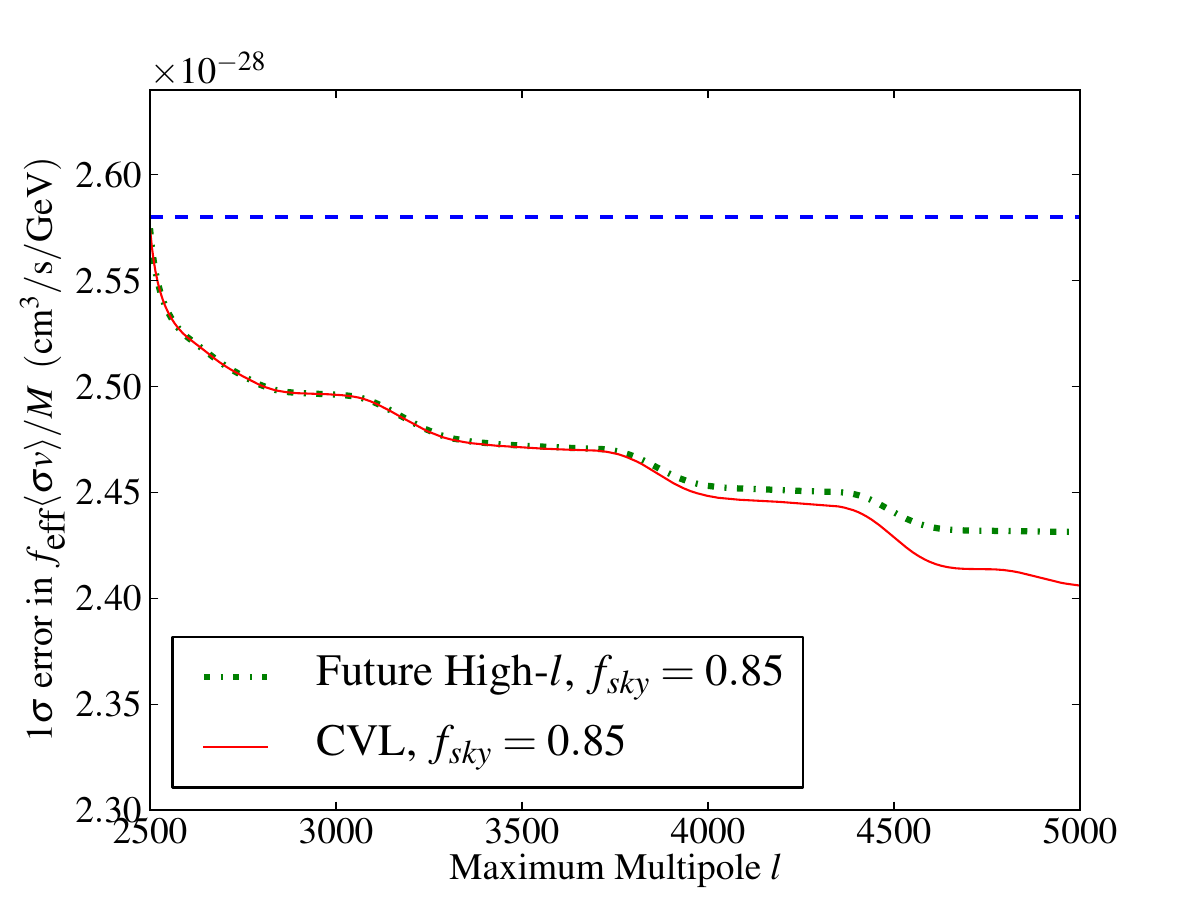}
     \caption{Fisher projected constraints including the complete Planck data from $2 < l < 2500$ (temperature and polarization) and extending it cumulatively for each multipole above $l=2500$ up to $l=5000$.  Experimental parameters are from a future high-$l$ experiment, and a cosmic variance limited experiment. The dashed line shows the Fisher projection for the full Planck temperature and polarization release (up to $l=2500$). The improvements over Planck are 6\% and 8\% respectively, including all $l$'s up to 5000.}
     \label{fig:total}
\end{figure}

The primary effects of dark matter annihilation on the CMB power spectra are an attenuation of power in both temperature and polarization especially at high-$l$, an enhancement of low-$l$ polarization power, and low-$l$ polarization peak shifts.  Since a number of cosmological parameters result in an attenuation of power at high-$l$ (e.g. $n_s$),  one would expect most of the constraining leverage on dark matter limits to come from the low-$l$ TE and EE spectra, which break parameter degeneracies.  To demonstrate the importance of low-$l$ polarization on improving constraints, we use Fisher forecasts to project the constraints obtainable by cumulatively adding the contribution to the Fisher matrix from each multipole below $l=500$ to the contribution from the range $500<l<5000$.  We use experimental parameters typical of Planck \cite{Planck}, a current generation polarization experiment like ACTpol, and a cosmic variance limited experiment (see Table \ref{tab:specs}). Including polarization information in the $100<l<500$ range improves the constraint by a factor of $\sim 3$ for ACTpol and $\sim 5$ for Planck (see Figure \ref{fig:lowl}).

In contrast, the constraint obtained from adding high-$l$ ($l>2500$) temperature and polarization spectra to the full Planck data (temperature and polarization, $2<l<2500$) plateaus around $l=4000$ for a future high-$l$ experiment (see Table \ref{tab:specs}), with no more than a 6\% improvement over full-Planck.  There is only an 8\% improvement over Planck for a cosmic variance limited experiment, including all $l$'s up to 5000 (see Figure \ref{fig:total}).

\begin{table}[t] 
\begin{threeparttable}
\caption{Experimental parameters used in forecasts}
\centering 
\begin{tabular}{c c c c c} 
\hline\hline 
 & Beam FWHM & $10^{6}\Delta T/T$ & $10^{6}\Delta T/T$ & $f_\text{sky}$ \\ [0.5ex] 
Experiment & (arcmin) & (I) & (Q,U) &  \\ [0.5ex] 
\hline 
Planck                           & 7.1 & 2.2 & 4.2 & 0.65  \\ 
ACTpol Ultrawide\footnote{We note that this represents just one possible configuration of the ACTpol survey.}                 & 1.4 & 4.5 & 6.3 & 0.24 \\ 
CMB Stage 4                      & 3.0 & 0.1 & 0.1 & 0.50 \\ 
Future High-$l$                  & 1.4 & 0.1 & 0.1 & 0.85  \\ 
\hline 
\end{tabular}
\label{tab:specs} 
\begin{tablenotes}
\small
\item Note: Noise values are indicated per beam.
\end{tablenotes}
\end{threeparttable}
\end{table}

\begin{figure}[t!]
     \begin{center}
      \subfigure{\includegraphics[width=.48\textwidth,natwidth=800,natheight=600]{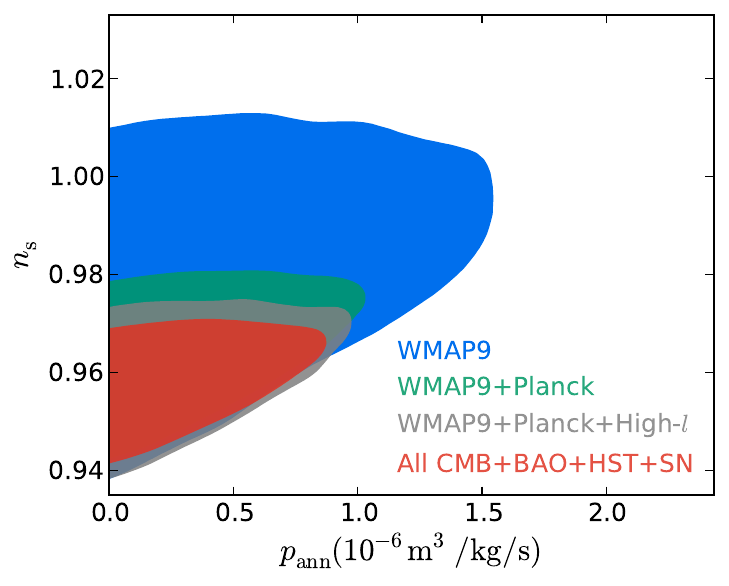}}
      \subfigure{\includegraphics[width=.48\textwidth,natwidth=800,natheight=600]{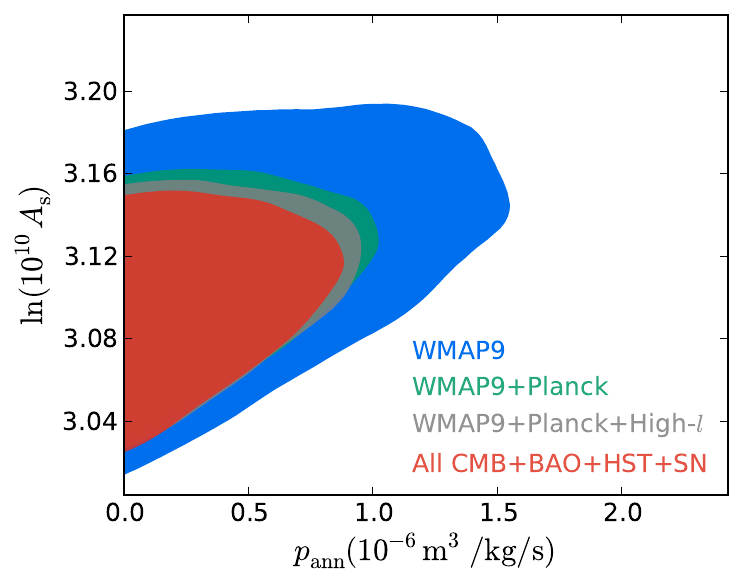}}
     \caption{95\% confidence limit contours for $n_s$ versus $p_\text{ann}$  and $\text{ln}(10^{10}A_s)$ versus $p_\text{ann}$, marginalized over the other parameters, for selected combinations of datasets.
     \vspace{-5mm}}
     \label{fig:contour}
     \end{center}
\end{figure}

\section{Current Constraints}\label{sec:constraints}

\begin{figure*}[th!]
     \begin{center}
      \includegraphics[width=.7\textwidth,natwidth=800,natheight=600]{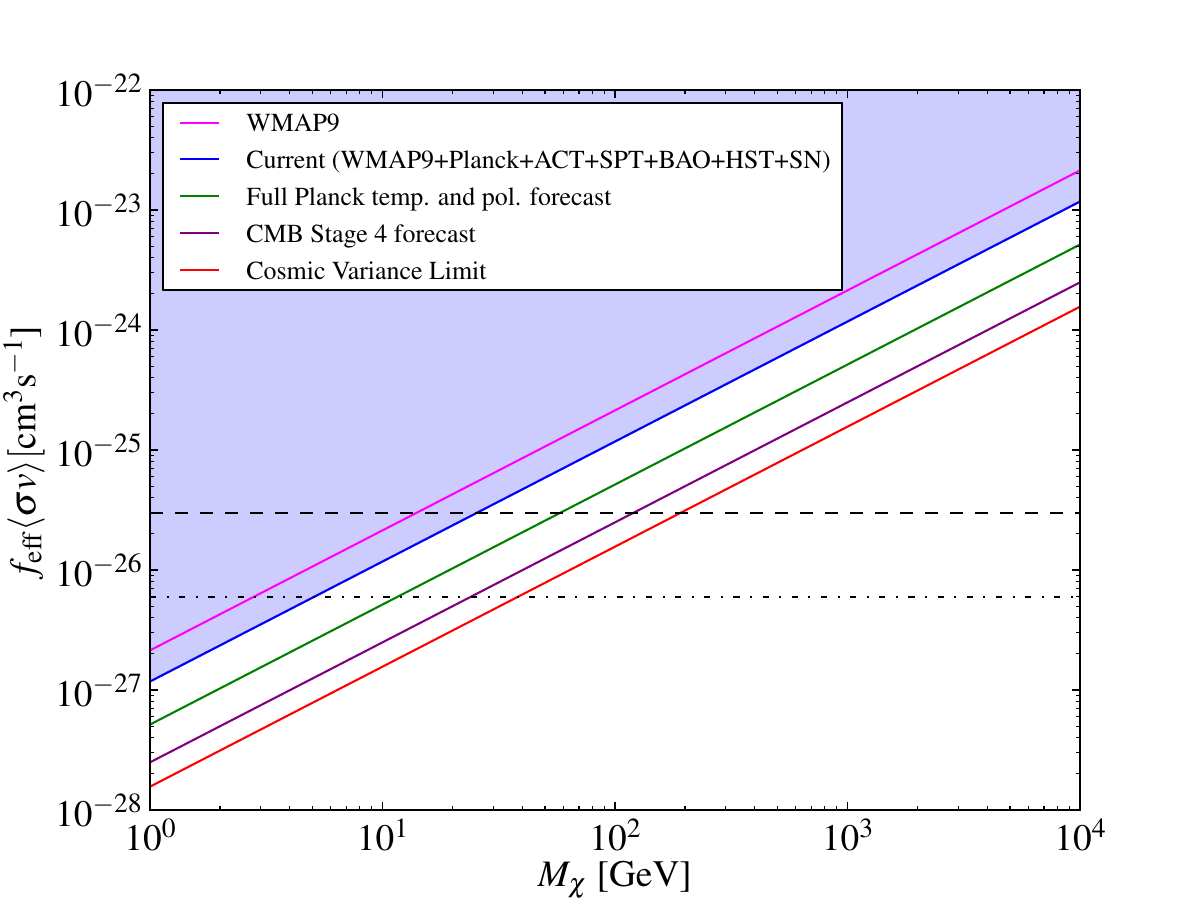}
     \caption{From top to bottom --- constraints on $p_{\text{ann}}$ from WMAP9 alone (pink) and from current data including WMAP9, Planck TT power spectrum and 4-point lensing signal, ACT, SPT, BAO, HST, and SN data (blue).  Also shown are Fisher forecasts for the complete Planck temperature and polarization power spectra (green), for a proposed CMB Stage IV experiment ($50<l<4000$ combined with $l<50$ from Planck, shown in purple), and for a cosmic variance limited experiment (up to $l=4000$) (red). The dashed line shows the thermal cross section of $3\times 10^{-26}\text{cm}^3\text{s}^{-1}$ for $f_\text{eff}=1$. The dot-dashed line shows the thermal cross section multiplied by a typical energy deposition fraction of $f_\text{eff}=0.2$ (see Table \ref{table:feffs}).
     \vspace{-5mm}}
     \label{fig:logplot}
     \end{center}
\end{figure*}

\begin{table*}[th!]
\caption{Upper limits at 95\% CL for $p_\text{ann}$ combining various datasets. The first column provides constraints when $p_\text{ann}$ is assumed to be constant with redshift. The second and third columns assume redshift-dependent energy deposition based on the `universal' curve discussed in Section \ref{sec:unisys}. The second column uses the original ``universal'' $e(z)$ curve derived in \cite{Finkbeiner2012}; the third column uses an updated curve that incorporates systematic corrections discussed in \cite{Galli2013}.}
\centering 
\begin{tabular}{c | c | c | c} 
\hline
Data Set &  Const. Ann. &  Non-Const.  Ann. &  Updated Non-Const. ($\text{m}^3\text{s}^{-1}\text{kg}^{-1}$)\\ [0.5ex] 
\hline
WMAP9 & $p_\text{ann}< 1.20 \times 10^{-6}$ & $p_\text{ann}< 1.26 \times 10^{-6}$ & $p_\text{ann}< 1.21 \times 10^{-6}$ \\
WMAP9 + Planck & $p_\text{ann}< 0.87 \times 10^{-6} $ & $p_\text{ann}< 0.85 \times 10^{-6}$ & $p_\text{ann}< 0.80 \times 10^{-6}$ \\
WMAP9 + Planck + Planck Lensing & $p_\text{ann}< 0.85 \times 10^{-6} $ & $p_\text{ann}< 0.86 \times 10^{-6}$ & $p_\text{ann}< 0.79 \times 10^{-6}$ \\
WMAP9 + Planck + Planck Lensing + ACT + SPT & $p_\text{ann}< 0.75 \times 10^{-6}$ & $p_\text{ann}< 0.75 \times 10^{-6}$ & $p_\text{ann}< 0.73 \times 10^{-6}$ \\
All CMB + BAO & $p_\text{ann}< 0.70 \times 10^{-6}$ & $p_\text{ann}< 0.66 \times 10^{-6}$ & $p_\text{ann}< 0.67 \times 10^{-6}$ \\
All CMB + BAO + HST & $p_\text{ann}< 0.71 \times 10^{-6}$ & $p_\text{ann}< 0.74 \times 10^{-6}$ & $p_\text{ann}< 0.66 \times 10^{-6}$ \\
All CMB + BAO + HST + Supernova & $p_\text{ann}< 0.70 \times 10^{-6}$ & $p_\text{ann}< 0.71 \times 10^{-6}$ & $p_\text{ann}< 0.66 \times 10^{-6}$ \\
\hline 
\hline
\end{tabular}
\label{table:constraints} 
\end{table*}

To obtain 95\% upper limits on $p_{\text{ann}}=f_\text{eff}\langle\sigma v\rangle/M_\chi$, we modified the recombination code \textsc{recfast} to include additional terms for the evolution of the hydrogen ionization fraction and matter temperature, given in Eqs. \ref{Eq:ionize} to \ref{Eq:temp}.  We performed a likelihood analysis on various datasets using the Markov Chain Monte Carlo code \textsc{cosmomc} \cite{Lewis2002}. We sampled the space spanned by $p_{\text{ann}}$ and the six cosmological parameters: $\Omega_bh^2$, $\Omega_ch^2$, $100\theta_*$, $\tau$, $n_s$, and $\mbox{ln}10^{10}A_s$.

Previous analyses using Planck data \cite{Ade2013} utilized only a small part of the WMAP9 polarization power spectrum \cite{Bennett2013}. Incorporating a larger range of the TE power spectrum can improve the constraint by up to a factor of $\sim 2.4$, depending upon how much of the WMAP9 polarization spectrum is included. Using Fisher forecasts, we find that the strongest constraint is obtained by including the WMAP9 TT+TE power spectrum from $l=2$ to $l=431$, and including the Planck TT spectrum for higher multipoles ($432< l <2500$). We also include `high-$l$' data -- a combination of ACT 2008-2010 \cite{Das2013} and SPT 2011-2012 \cite{Schaffer2011} observations, using their power spectra in the range $2500<l<4500$, which is included in the publicly available Planck likelihood \cite{Ade2013b}. Several low-redshift (non-CMB) datasets are also combined. These include baryon acoustic oscillation data (BAO) from BOSS DR9 \cite{Dawson2013}, Hubble Space Telescope measurements of over 600 Cepheid variables (HST) \cite{Riess2011}, and supernovae type Ia data from the Union 2.1 compilation (SN) \cite{Suzuki2012}.

When combining CMB datasets, we do not account for the covariance between disjoint $l$-ranges from different experiments as we expect this to be negligible \cite{Ade2013}.  In using the Planck likelihood code, we removed the TT power spectrum contribution from $l<431$ by setting the relevant diagonal elements of the covariance matrix to effectively infinity ($10^{10}$) and the off-diagonal elements to zero.\footnote{We note that there is a $2.49\%$ calibration difference between the Planck and WMAP9 power spectra \cite{Ade2013}.  Since the origin of this offset is unclear, in this work we take each dataset as given and do not adjust either.}

The dark matter annihilation constraints thus obtained are listed in Table \ref{table:constraints}. We checked for convergence of the chains using a Gelman-Rubin test statistic, ensuring that the corresponding $R-1$ fell below 0.01. We obtained three sets of constraints, one with constant $p_\text{ann}$, one with $p_\text{ann}(z)$ proportional to the original universal $e(z)$ curve (shown as the blue curve in Figure \ref{fig:ewimp}) to account for a generic redshift dependence of the energy deposition, and one with $p_\text{ann}(z)$ proportional to an updated universal $e(z)$ curve that includes systematic corrections as detailed in Section \ref{sec:unisys}.  The constraints using the updated universal curve with systematic corrections are also shown in Figure \ref{fig:logplot}. In general, there is a small improvement in the constraints using the updated $e(z)$ curve incorporating systematic corrections. As discussed above, this is not expected a priori from the Fisher matrix analysis using the CMB data only; it likely reflects some combination of the breakdown of the approximations in the Fisher matrix approach, differences between the data and the idealized $\Lambda$CDM baseline used for the Fisher analysis, the effect of including non-CMB datasets, and the few-percent uncertainty in the constraints due simply to scatter between CosmoMC runs. 

\begin{figure}[ht!]
     \begin{center}
      \includegraphics[width=.53\textwidth,natwidth=800,natheight=600]{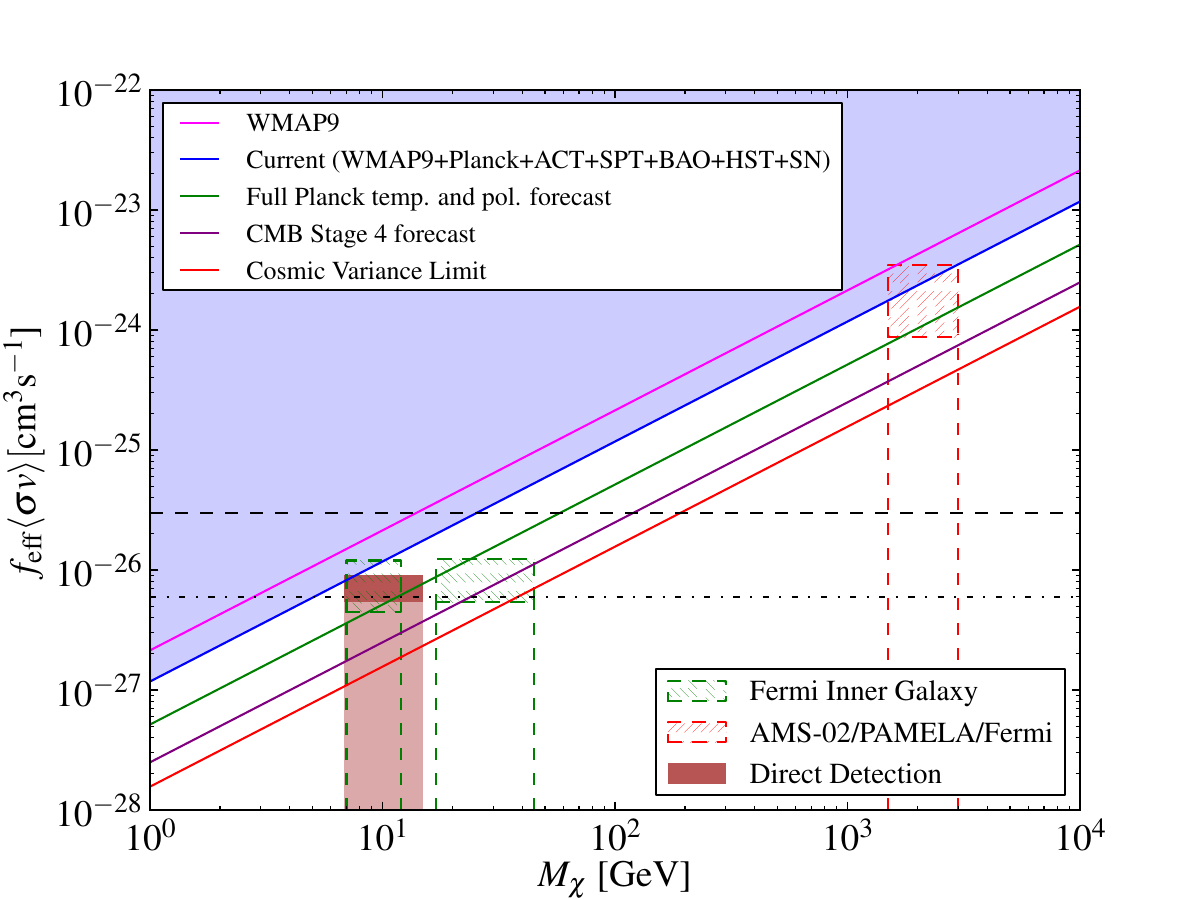}
     \caption{Current constraints are compared with dark matter model fits to data from other indirect and direct dark matter searches. The data from indirect searches include that from AMS-02, PAMELA, and Fermi, and the data from direct searches include that from CDMS, CoGeNT, CRESST, and DAMA. The lighter shaded direct detection region allows for p-wave annihilations, and the dashed vertical lines for the indirect detection regions allow for p-wave annihilations for non-thermally produced dark matter.
     \vspace{-6mm}}
     \label{fig:logplotmodels}
     \end{center}     
\end{figure}

The greatest improvement to the WMAP9-only constraint comes from adding the Planck TT spectrum ($\sim 50\%$) as it particularly constrains the spectral index $n_s$ which is strongly degenerate with the annihilation parameter $p_\text{ann}$ (see Figure \ref{fig:contour}). The high-$l$ CMB and BAO datasets improve our constraints by 8\% and 9\%, respectively.  Adding to this the HST and Supernova data do not considerably improve these limits.

\section{Discussion}\label{sec:models}

The constraint obtained from using the updated universal deposition curve and including all available datasets is a factor of $\sim 2$ stronger than that from WMAP9 data alone \cite{Ade2013}. The strongest constraint, including all available data, of  $p_\text{ann}< 0.66 \times 10^{-6} \text{m}^3\text{s}^{-1}\text{kg}^{-1}$ at $95\%$ CL, excludes annihilating dark matter of masses $M_\chi<26$ GeV, assuming a thermal cross section of $3\times 10^{-26} \text{cm}^3\text{s}^{-1}$ and perfect absorption of injected energy ($f_\mathrm{eff}=1$). Using a more realistic absorption efficiency of $f_\mathrm{eff}=0.2$, we exclude annihilating thermal dark matter of masses $M_\chi<5$ GeV at the $2\sigma$ level.\footnote{This constraint on $p_\text{ann}$ is a factor of two weaker than that found by \cite{LopezHonorez2013}, possibly due to the priors chosen in that work.}

These constraints can be compared to dark matter models explaining a number of recent anomalous results from other indirect and direct dark matter searches.  
Recent measurements by the AMS-02 collaboration \cite{Aguilar2013} confirm a rise in the cosmic ray positron fraction at energies above 10 GeV, which was found earlier by the PAMELA \cite{Adriani2013} and Fermi collaborations \cite{Ackermann2012}. Such a rise is not easy to reconcile with known astrophysical processes, although contributions from Milky Way pulsars within $\sim1$ kpc of the Earth could provide a possible explanation \cite{Hooper2009,Yuksel2009,Profumo2012,Malyshev2009,Grasso2009}.  Dark matter annihilating within the galactic halo also remains a possible explanation of the positron excess \cite{Jin2013,Cholis2013,Bergstrom2013,Kopp2013}. Dark matter models considered in \cite{Cholis2013} to explain the AMS-02/PAMELA positron excess cannot have significant annihilation into Standard Model gauge bosons or quarks in order to be consistent with the antiproton-to-proton ratio measured by PAMELA, which is found to agree with expectations from known astrophysical sources \cite{Adriani2010}. In addition, the combination of the Fermi electron plus positron fraction \cite{Abdo2009,Ackermann2010} and the AMS-02/PAMELA positron excess suggest that a viable dark matter candidate would need to have a mass greater than $\sim1$~TeV.  As found by \cite{Cholis2013}, dark matter particles in the $\sim 1.5-3$ TeV range with a cross section of $\langle\sigma v\rangle\sim(6-23)\times10^{-24} \text{cm}^3/\text{s}$, that annihilate into light intermediate states that in turn decay into muons and charged pions, can fit the Fermi, PAMELA, and AMS-02 data. Direct annihilations into leptons do not provide good fits \cite{Cholis2013}. Such high cross sections can be reconciled with the current dark matter abundance in the Universe in three ways: 
\begin{inparaenum}[(i)] \item Dark matter can have a thermal cross section at freeze-out, and the cross section can have a $1/v$ dependence, called Sommerfeld enhancement \cite{Pospelov2008,ArkaniHamed2009}.  If the cross section is Sommerfeld enhanced to be $\sim10^{-24}$ today in the Galactic halo, then it would be orders of magnitude larger at recombination (since $v_{\mathrm{recom}}<v_{\mathrm{halo}}$).  Such a possibility is strongly excluded by the CMB constraints (as noted in \cite{Slatyer2009}) for a wide range of masses including those that fit the AMS-02 data.
\item Dark matter has a thermal cross section at freeze-out, and Sommerfeld enhancement saturates at a cross section of $\sim 10^{-24} \text{cm}^3/\text{s}$.  So dark matter has this cross section just before (and during) recombination, and also in the halo of the Milky Way. 
\item Dark matter particles are non-thermal, in which case the cross section has always been ($\sim 10^{-24} \text{cm}^3/\text{s}$).
\end{inparaenum}
The last two possibilities are shown in Figure \ref{fig:logplotmodels}, and are probed but not excluded by our current constraints. Here we use the updated $f_\text{eff}$ values from Table \ref{table:feffs} corresponding to the best-fit annihilation channels found by \cite{Cholis2013}.  

\begin{table}[b!]
\vskip -4mm
\caption{Effective energy deposition fractions for 41 dark matter models. The third column is an updated version of Table I in \cite{Slatyer2009}, and the fourth column includes systematic corrections discussed in Section \ref{sec:unisys}.\\}
\centering 

\resizebox{0.75\columnwidth}{!}{%

\begin{tabular}{c c c c} 
\hline\hline 
Channel & DM Mass (GeV) & $f_{\text{eff}}$ & $f_{\text{eff,new}}$ \\ [0.5ex] 
\hline 
Electrons                           & 1 & 0.85 & 0.45  \\ 
$\chi\chi\rightarrow e^{+}e^{-}$     & 10 & 0.77 & 0.67  \\ 
                                    & 100 & 0.60 & 0.46  \\ 
                                    & 700 & 0.58 & 0.45  \\ 
                                    & 1000 & 0.58 & 0.45  \\ 
\hline 
Muons                                 & 1 & 0.30 & 0.21\\ 
$\chi\chi\rightarrow \mu^{+}\mu^{-}$     & 10 & 0.29 &  0.23\\ 
                                    & 100 & 0.23 & 0.18\\ 
                                    & 250 & 0.21 & 0.16\\ 
                                    & 1000 & 0.20 & 0.16\\ 
                                    & 1500 & 0.20 & 0.16\\ 
\hline 
Taus                                 & 200 & 0.19 & 0.15\\ 
$\chi\chi\rightarrow \tau^{+}\tau^{-}$ & 1000 & 0.19 & 0.15\\ 
\hline 
XDM electrons                        & 1 & 0.85 & 0.52\\ 
$\chi\chi\rightarrow \phi\phi$     & 10 & 0.81 & 0.67\\ 
followed by                         & 100 & 0.64 & 0.49\\ 
$\phi\rightarrow e^{+}e^{-}$          & 150 & 0.61 & 0.47\\ 
                                     & 1000 & 0.58 & 0.45\\ 
\hline 
XDM muons                          & 10 & 0.30 & 0.21\\ 
$\chi\chi\rightarrow \phi\phi$     & 100 & 0.24 & 0.19\\ 
followed by                         & 400 & 0.21 & 0.17\\ 
$\phi\rightarrow \mu^{+}\mu^{-}$      & 1000 & 0.20 & 0.16\\ 
                                    & 2500 & 0.20 & 0.16\\ 
\hline 
XDM taus                                 & 200 & 0.19 & 0.15\\ 
$\chi\chi\rightarrow \phi\phi, \phi\rightarrow\tau^{+}\tau^{-}$ & 1000 & 0.18 & 0.14\\ 
\hline 
XDM pions                          & 100 & 0.20 & 0.16\\ 
$\chi\chi\rightarrow \phi\phi$     & 200 & 0.18 & 0.14\\ 
followed by                        & 1000 & 0.16 & 0.13\\ 
$\phi\rightarrow \pi^{+}\pi^{-}$     & 1500 & 0.16 & 0.13\\ 
                                    & 2500 & 0.16 & 0.13\\ 
\hline 
W bosons                          & 200 & 0.26 & 0.19\\ 
$\chi\chi\rightarrow W^{+}W^{-}$     & 300 & 0.25 & 0.19\\ 
                                   & 1000 & 0.24 & 0.19\\ 
\hline 
Z bosons                          & 200 & 0.24 & 0.18\\ 
$\chi\chi\rightarrow ZZ$         & 1000 & 0.23 & 0.18\\ 

\hline 
Higgs bosons                          & 200 & 0.30 & 0.22\\ 
$\chi\chi\rightarrow h\bar{h}$         & 1000 & 0.28 & 0.22\\ 

\hline 
b quarks                               & 200 & 0.31 & 0.23\\ 
$\chi\chi\rightarrow b\bar{b}$         & 1000 & 0.28 & 0.22\\ 

\hline 
Light quarks                               & 200 & 0.29 & 0.22\\ 
$\chi\chi\rightarrow u\bar{u}, d\bar{d} \mbox{   (50\% each)}$         & 1000 & 0.28 & 0.21\\ 
\hline 
\end{tabular}
}

\label{table:feffs} 
\end{table}

One additional possibility is that dark matter has a p-wave annihilation cross section with a $\sim v^2$ dependence on velocity. Dark matter that has a p-wave cross section and fits the AMS-02/PAMELA data would have to be non-thermal, since the cross section during freezeout would be orders of magnitude larger and would vastly over-deplete the relic density.  Since $v_{\mathrm{recom}}\ll v_{\mathrm{halo}}$, the cross section around recombination can be orders of magnitude smaller in this case.  We indicate this by dashed vertical lines in Figure \ref{fig:logplotmodels}.

Recent direct detection experiments such as CDMS, CoGeNT, CRESST, and DAMA, have also reported anomalous signals that could potentially be interpreted as arising from dark matter \cite{Bernabei2013,Agnese2013,Aalseth2011,Angloher2012}.  For example, the CDMS collaboration recently reported three events above background where they expected only 0.7 events, by measuring nuclear recoils using Silicon semiconductor detectors operating at 40 mK \cite{Agnese2013}. If the CDMS anomalous events are explained by dark matter, then they favor a best-fit dark matter mass of $8.6$ GeV and a dark matter-nucleon cross section of $1.9\times 10^{-41}\text{cm}^2$ (with 68\% CL ranges of 6.5-15 GeV and $2\times 10^{-42}-2\times 10^{-40} \text{cm}^2$) (see Figure 4 in \cite{Agnese2013}). The dark matter candidates that potentially explain the anomalous signals from the other direct detection experiments have best-fit regions that do not completely overlap in the two-dimensional mass/nucleon cross section space, but have mass ranges that are comparable \cite{Agnese2013}.  If we assume a thermal s-wave annihilation cross section during the recombination era and an $f_\text{eff}$ from Table \ref{table:feffs} corresponding to annihilation into $b\bar{b}$, the current constraints presented above start to probe, but do not exclude, such a dark matter candidate.  However, future Planck results and those from a proposed CMB Stage IV experiment \cite{Abazajian2013,Abazajian2013b} will more definitively probe the relevant regime, as shown in Figure \ref{fig:logplotmodels}.  If dark matter has p-wave annihilations instead, then generic thermal dark matter can have annihilation cross sections at recombination orders of magnitude lower than the thermal cross section.  This is indicated by a lighter shaded direct detection region in Figure \ref{fig:logplotmodels}.

Observations of the Galactic Center and inner Galaxy by the Fermi Gamma-ray Telescope reveal an extended Gamma-ray excess above known backgrounds, peaking at around 2-3 GeV.  A population of unresolved millisecond pulsars has been proposed as a possible explanation, but as found by \cite{Hooper2013}, in order for pulsars to reproduce the excess in the inner Galaxy their luminosities and abundances would need to be quite different from any observed pulsar population.
However, these measurements are well fit by dark matter particles with mass in the ranges 7-12 GeV (if annihilating mostly to leptons) and 25-45 GeV (if annihilating mostly to hadrons), and are consistent with a cross section of $\sim 10^{-26} \text{cm}^3/\text{s}$ \cite{Hooper2011,HooperFermi2011, HooperFermi2013, Gordon2013}.  For the higher mass range, we assume annihilations into quarks and gauge bosons and a thermal cross section.  For the lower mass range, we assume annihilations into muons and taus and a thermal cross section.  Figure \ref{fig:logplotmodels} shows that we can probe but not exclude this interpretation. The complete Planck data will better examine this possibility, as will data from the proposed CMB Stage IV experiment. 

The constraints on dark matter annihilation cross section and mass from the CMB are complementary and competitive with other indirect detection probes, and offer a relatively clean way to measure dark matter properties in the early Universe.  Current CMB experiments are starting to probe very interesting regions of dark matter parameter space, and future CMB polarization measurements have the potential to significantly expand the constrained regions or detect a dark matter signal. 

 \vspace{2mm}
\acknowledgments
 \vspace{-2mm}
 
The authors thank Erminia Calabrese and Silvia Galli for very useful correspondence, and especially Ren\`{e}e Hlozek for help with the Planck likelihood. The authors also acknowledge helpful discussions with Alexander van Engelen, Rouven Essig, and Neal Weiner.  M.M. is supported by an SBU-BNL Research Initiatives Seed Grant: Award Number 37298, Project Number 1111593. This work is supported by the U.S. Department of Energy under cooperative research agreement Contract Number DE-FG02-05ER41360. The authors also gratefully acknowledge the use of the software packages CAMB and CosmoMC, and the publicly available Planck and WMAP likelihoods. Some computations were performed on the GPC supercomputer at the SciNet HPC Consortium. SciNet is funded by: the Canada Foundation for Innovation under the auspices of Compute Canada; the Government of Ontario; Ontario Research Fund -- Research Excellence; and the University of Toronto.

\bibliography{refs}

\end{document}